\documentclass[prc,floatfix,nofootinbib,showpacs,showkeys,preprint]{revtex4-2}
\usepackage[utf8]{inputenc}
\usepackage[T1]{fontenc}
\usepackage{amsmath}
\usepackage{amsfonts}
\usepackage{amssymb}
\usepackage{graphicx}

\begin{document}
	\title{Neutrinos at FPF}
	\author{Ulrich Mosel}
	\email[Contact e-mail:]{mosel@physik.uni-giessen.de}
	\author{Kai Gallmeister}
	\affiliation{Institut f\"ur Theoretische Physik, Universit\"at Giessen, Giessen, Germany}
	
	\begin{abstract}
		The GiBUU model is used to obtain information on possible neutrino-nucleus events at the proposed Forward Physics Facility (FPF) at CERN. An FPF neutrino program could contribute to fundamental questions such as formation times, color transparency and the EMC effect for neutrinos.
	\end{abstract}
	
	\maketitle

\section{GiBUU as Generator}
The Giessen model is an extensive general theory framework to describe nuclear collisions, from relativistic heavy-ion collisions to neutrino-induced reactions on nuclear targets.  For its treatment of final state interactions (FSI) it is built on quantum-kinetic transport theory \cite{Kad-Baym:1962} and as such can be used as a generator, called GiBUU, for the full final state of a reaction. The underlying theory is described in detail in \cite{Buss:2011mx}; this article also contains the details of the actual implementation of theory and its parameters in the code. The code is being updated from year to year and its source-code is freely available from \cite{gibuu}.

In the following section some results obtained with GiBUU for neutrinos with an incoming energy of 1 TeV impinging on a Tungsten target are discussed. At this energy Deep Inelastic Scattering (DIS), handled by PYTHIA inside GiBUU, is dominant.

\section{Inclusive Properties of 1 TeV muon neutrinos on a W target}
\begin{figure}[hb]
	\centering
	\includegraphics[width=0.7\linewidth]{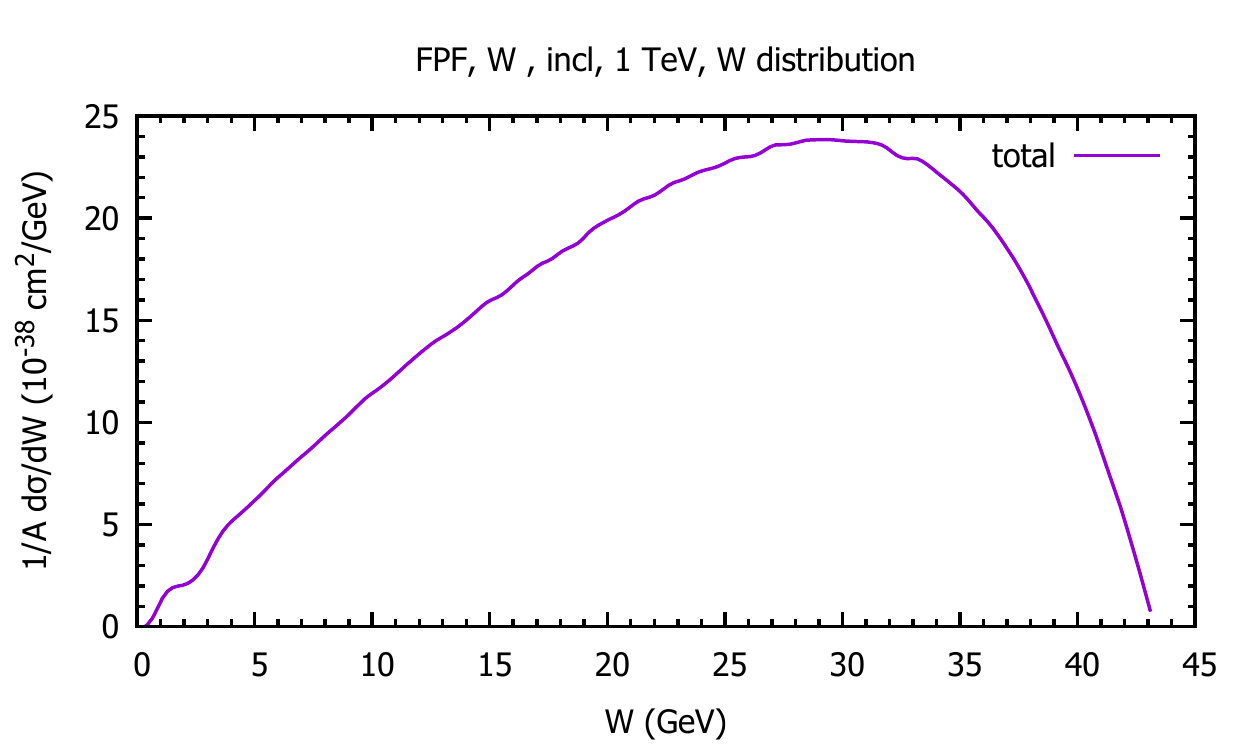}
	\caption{Distribution of invariant mass populated in a 1 TeV neutrino reaction on a Tungsten target}
	\label{fig:fpf-w}
\end{figure}

\begin{figure}
		\centering
		\includegraphics[width=0.7\linewidth]{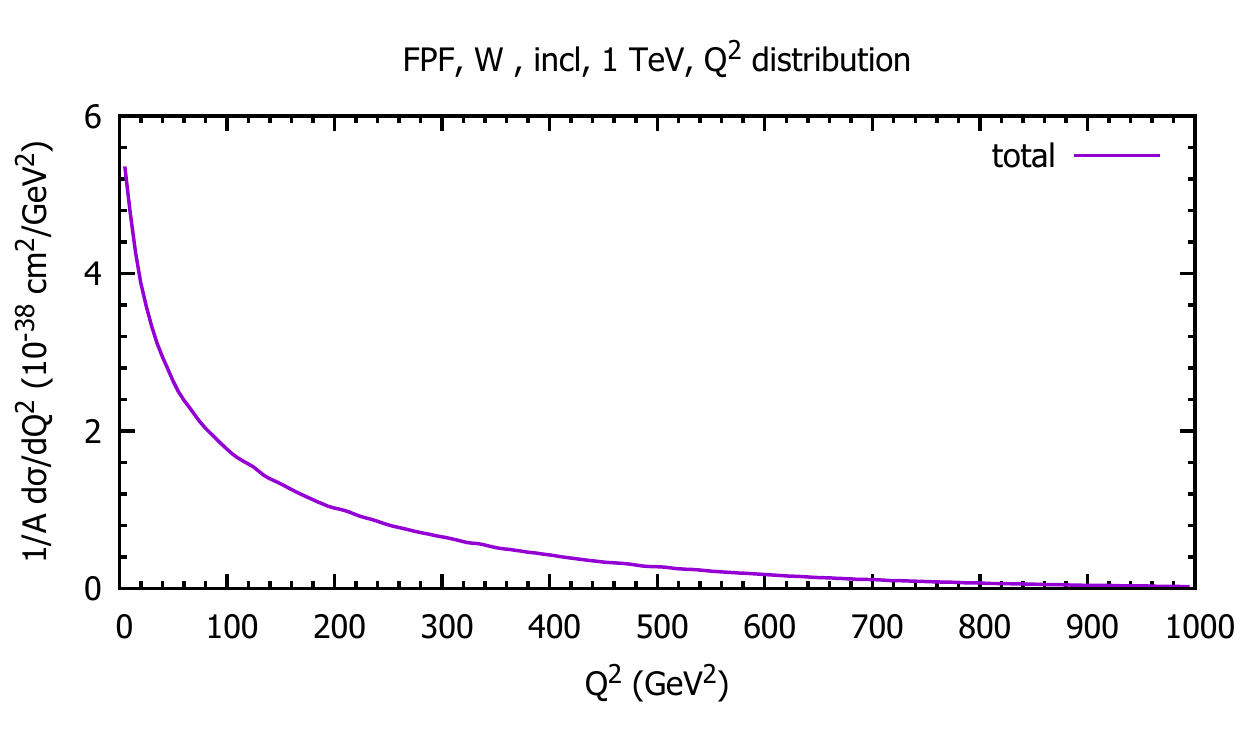}
		\caption{$Q^2$ distribution}
		\label{fig:fpf-q2}
	\end{figure}

In order to get a first impression of the kinematical regimes accessible by this reaction we show in Fig.\ \ref{fig:fpf-w} the inclusive invariant mass distribution. The distribution peaks at about $W = 30$ GeV, i.e.\ well in the DIS regime.
This is also reflected in the four-momentum transfer distribution given in Fig.\ \ref{fig:fpf-q2} which reaches up to very high values.
The energy-transfer ($\omega$) distribution, on the other hand, is fairly flat with a value (per nucleon) decreasing from about $0.8 \cdot 10^{-38}$ at low $\omega$ to about $0.6 \cdot 10^{-38}$ cm$^2$/GeV at 1 TeV.

\begin{figure}[hbt]
		\centering
		\includegraphics[width=0.7\linewidth]{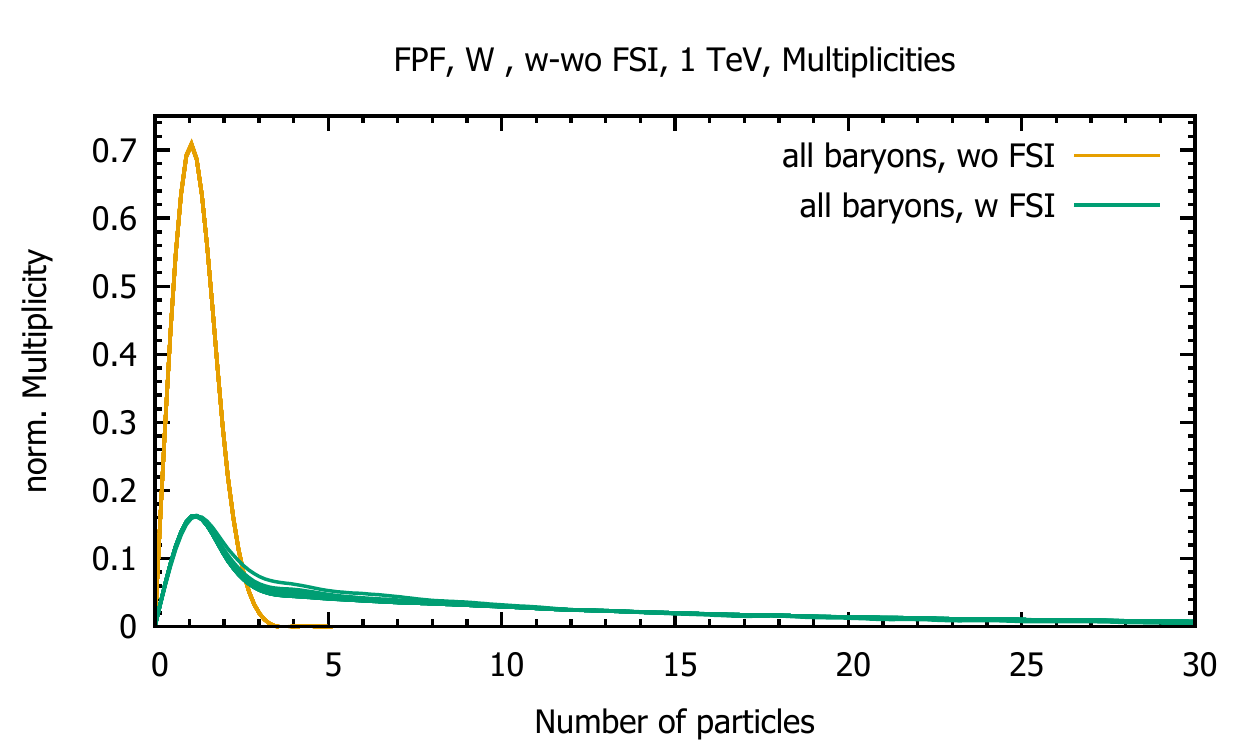}
		\caption{Multiplicity distribution of baryons without and with FSI}
		\label{fig:fpf-multiplfsi-compallbaryons}
	\end{figure}

We now focus on the multiplicities of final-state particles. Fig.\ \ref{fig:fpf-multiplfsi-compallbaryons} shows the multiplicity distribution of final-state baryons. The top (yellow) curve gives the result before any final state interactions. This distribution peaks at multiplicity = 1 with a tail up to about 3 - 4. This tail is caused by the production of baryon-antibaryon pairs. When FSI are turned on the multiplicity distribution changes significantly: the peak height is decreased by about a factor of 5 and a long tail reaching up to about 25 develops. This is a consequence of the so-called 'avalanche effect' in which initial nucleons collide with others on the way out of the target.	
 
\begin{figure}[hbt]
		\centering
		\includegraphics[width=0.7\linewidth]{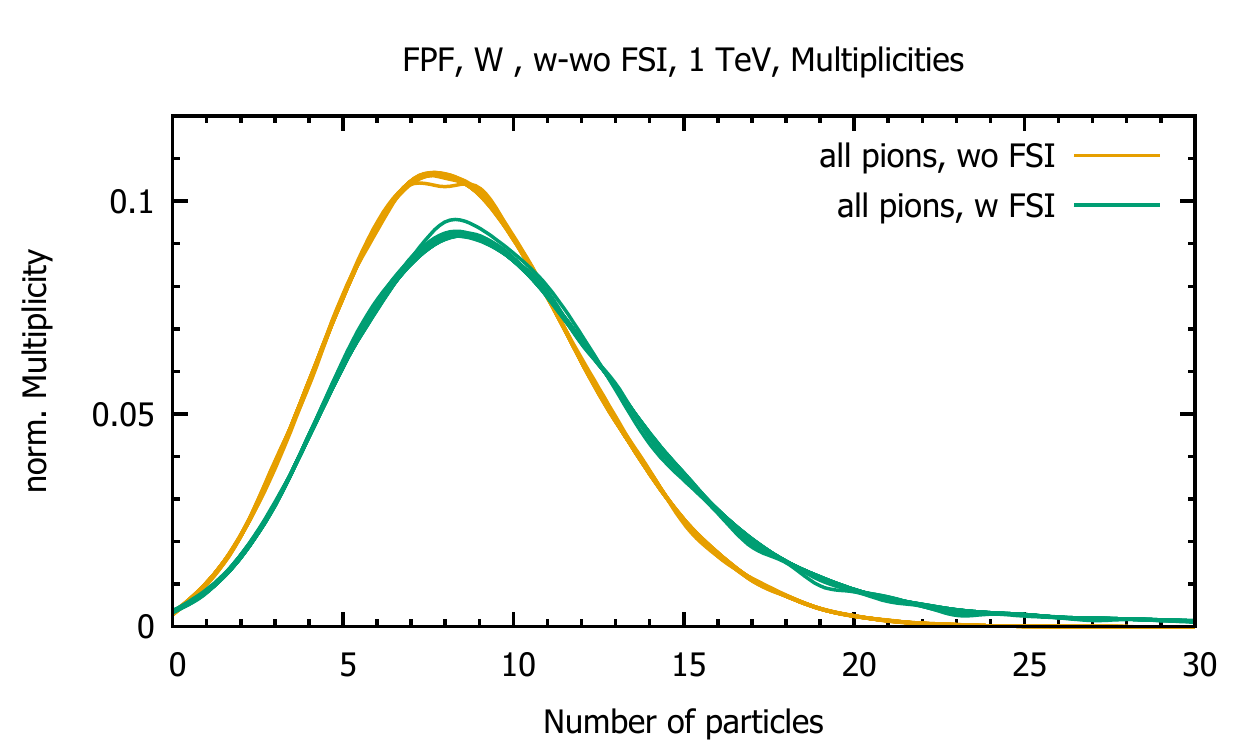}
		\caption{Multiplicity distribution of pions without and with FSI}
		\label{fig:fpf-multiplfsi-compallpions}
	\end{figure}

 The pion multiplicities, on the other hand show a much less effect of FSI (see Fig.\ \ref{fig:fpf-multiplfsi-compallpions}). The initial pion multiplicity distribution hardly changes.

\begin{figure}[h]
		\centering
		\includegraphics[width=0.7\linewidth]{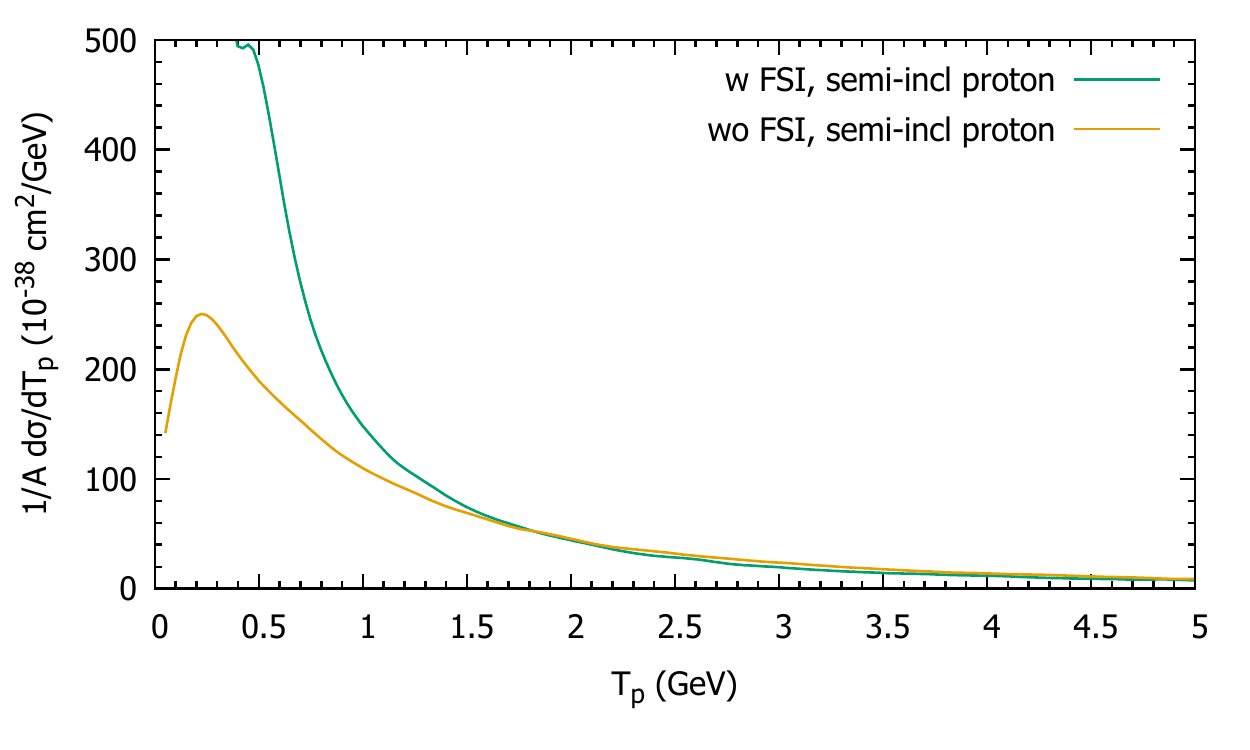}
		\caption{FSI effects on proton spectra}
		\label{fig:fpf-spectraexcl-incl}
	\end{figure}

The baryon avalanche effect also shows up in the kinetic energy distribution (Fig.\ \ref{fig:fpf-spectraexcl-incl}). While the events before FSI show a peak at about 0.3 GeV, the final state interactions change that distribution significantly. Cross sections are decreased at high $T_p > 2$ GeV and are dramatically increased at lower $T_p < 1.5$ GeV. This is again a consequence of the avalanche effect: initial high energy baryons collide with target nucleons, thus  increasing the multiplicity and consequently loosing energy.

The effect of FSI on the spectra of baryons is obviously quite significant. This opens the possibility to study phenomena such as Color Transparency (CT) and hadron formation in medium \cite{Gallmeister:2007an}.

\section{Conclusion}
At 1 TeV bombarding energy the DIS process is dominant. Any science program exploiting these neutrinos at the FPF should thus concentrate on this reaction mechanism. Such experiments could yield very valuable information on
\begin{itemize}
	\item {\bf Formation times in DIS events} Earlier analyses based on HERMES and EMC electron data had shown that only hadronic FSI cross sections that increase linearly with time can describe both data sets simultaneously \cite{Gallmeister:2007an}. Data at the FPF could help to validate this result in a new kinematical regime.
	
	\item {\bf Color Transparency (CT) effects} The very recent, unexpected result from JLAB that CT for baryons does not set in up to $Q^2 = 14$ GeV$^2$ presents a challenge to standard CT theory \cite{HallC:2020ijh}. It has been argued that only final state baryons that originate in a DIS event should be subject to CT \cite{Mosel:2021CT}. Experiments at the FPF where all events are dominated by DIS are ideal to investigate this further. 
	
	\item {\bf Tension in the EMC effect for electrons and neutrinos} It has been known for some time that the EMC effect seems to be different for electrons and neutrinos; the latter do not seem to show the strong antishadowing effect seen with electrons \cite{SajjadAthar:2020nvy}. This result presents a challenge to pQCD and thus calls  for further verification in a new kinematical regime.
\end{itemize}

\begin{acknowledgments} 
	This work has been partially supported by the HFHF.
\end{acknowledgments}	

\bibliography{nuclear.bib}
\end{document}